\documentclass[aps,pra,twocolumn,amsmath,amssymb,superscriptaddress,floatfix]{revtex4-1}

\usepackage{graphicx}
\usepackage{multirow}
\usepackage{color}
\usepackage{bm}

\def\t#1{\textrm{#1}}
\def\ket#1{|#1\rangle }
\def\bra#1{\langle #1 |}
\def\braket#1{\langle #1 \rangle}
\def\n{\nonumber \\ }

\begin{document}

\title{Shift current from electromagnon excitations in multiferroics}

\author{Takahiro~Morimoto}
\affiliation{Department of Applied Physics, The University of Tokyo, Hongo, Tokyo, 113-8656, Japan}
\affiliation{JST, PRESTO, Kawaguchi, Saitama, 332-0012, Japan}

\author{Naoto~Nagaosa}
\affiliation{Department of Applied Physics, The University of Tokyo, Hongo, Tokyo, 113-8656, Japan}
\affiliation{RIKEN Center for Emergent Matter Sciences (CEMS), Wako, Saitama, 351-0198, Japan}

\begin{abstract}
Electromagnon is the spin wave in multiferroic materials
and is known to accompany electric polarization due to the cross correlation between the charge and spin.
Here, we theoretically show that the electromagnons also induce dc current upon their photoexcitations.
The proposed dc current response originates from the shift current mechanism which is characterized by the so called shift vector, a geometric quantity of the Bloch wavefunctions.
\end{abstract}

\date{\today}

\maketitle

\section{Introduction}
 Multiferroics, i.e., the coexistence of the multiple 
broken symmetries, is a subject of intensive studies 
with the focus on the coupled dynamics of the 
multiple order parameters. Most of the works are
on the magnetic ferroelectrics, i.e.,
the time-reversal $\mathcal{T}$ and inversion $\mathcal{P}$ symmetries
are broken simultaneously, which enables the 
enhanced magnetoelectric (ME) effect
and electrical (magnetic) control of the spins 
(electric polarization)~\cite{Kimura03,Katsura05,Mostovoy06,Fiebig05,Tokura14}. 
The elementary excitation of multiferroics is
electromagnon, i.e., the spin wave, which 
is accompanied by the fluctuation of the 
electric polarization, and hence can be excited
by the electric field. The THz and infrared spectroscopy 
of electromagnon has been intensively
studied and also its giant nonreciprocal optical effect
has been discovered~\cite{Kibayashi14}.  

  On the other hand, nonreciprocal phenomena in 
noncentrosymmetric quantum materials begin to 
attract increasing recent attentions, where the 
$\mathcal{P}$ and $\mathcal{T}$ symmetries play essential roles~\cite{Tokura18}.  
The broken $\mathcal{P}$ and/or $\mathcal{T}$ symmetries 
are encoded by the Berry phase of the 
Bloch wavefunctions of electrons in crystals.
One novel consequence of this Berry phase is
the shift current, i.e.,
the photocurrent induced by the interband 
transition from light excitation even when there is no external
dc electric field~\cite{Sipe,Young-Rappe,Morimoto-Nagaosa16,Nagaosa-Morimoto17,Cook17,Sotome18}. 
This phenomenon originates from 
the difference in the intracell coordinates
between the conduction and valence bands (i.e. so called shift vector), which is described by the Berry connection difference between the two bands.
When the interband transition occurs, the 
change in the intracell coordinates results in the
current, and continuous excitation in the 
steady state results in the dc photocurrent.

This shift current is distinct from the conventional 
photocurrent by the photoexcited charge carriers,
and remarkably expected to be 
induced even when the exciton is photoexcited below the band gap~\cite{Morimoto-exciton16,Chan19}.
Note that the shift current does not require the 
broken $\mathcal{T}$ symmetry; only $\mathcal{P}$ symmetry breaking
is enough for shift current generation as far as it is not forbidden by other crystalline symmetries.
In the absence of the spin-orbit interaction,
light irradiation can only excite the excitons in the spin singlet states, and hence, its implication to magnetic states has not been discussed so far.

In the present paper, we theoretically study the
shift current induced by the electromagnon. 
We do so by combining and unifying the two streams of studies mentioned above, i.e., the multiferroics and 
the shift current. The basic idea is that 
the spin wave can be regarded as the 
triplet excitons, the electron-hole pairs in the spin triplet states.
The triplet exciton acquires an oscillator strength in the light-matter coupling when the spin-orbit interaction is present.
Therefore, we can expect that the electromagnon excitation by light supports the dc shift current, in addition to the electric polarization (Fig.~\ref{fig: model}(a)).
In the rest of this paper, we show that indeed this scenario is true by demonstrating the shift current response in a one dimensional toy model of multiferroics.

\section{1D model of a multiferroic ferromagnet}
\begin{figure*}
\begin{center}
\includegraphics[width=0.8\linewidth]{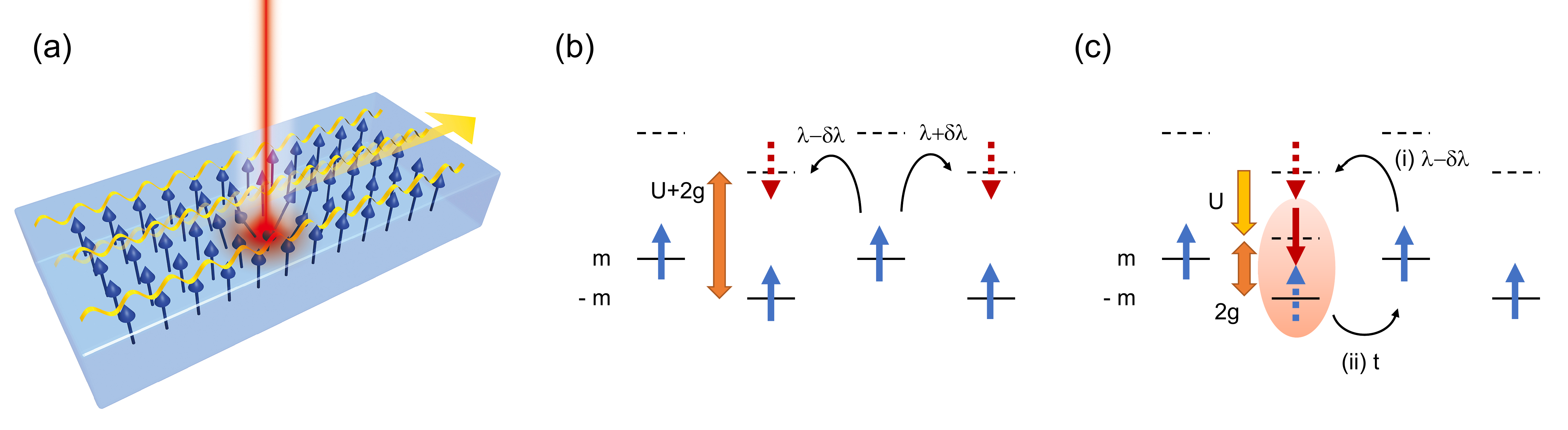}
\caption{\label{fig: model}
(a) Schematic picture of the shift current from electromagnon excitations.
(b) Schematic picture of the 1D model. In the ground state, the electron number is 1 per site, and the strong Coulomb interaction $U$ induces the magnetic ordering, which is assumed to be ferromagnetic stabilized by the spin anisotropy $g$.
The alternating spin orbit interaction $\lambda \pm \delta \lambda$ breaks the inversion symmetry $\mathcal P$.
(c) Electromagnon excitation in the 1D model. The electromagnon can be regarded as a triplet exciton state which is a bound state of an electron with down spin and a hole with up spin.
The electromagnon acquires optical activity due to the spin orbit coupling. In the 1D model, the electromagnon is formed, for example, by the process where (i) the photoexcitation through spin orbit coupling creates an electron hole pair in the triplet state in the neighboring sites, and then (ii) an electron or hole move to the same site through the hopping term $t$. The triplet exciton state is stabilized by an effective interaction $U$ between the electron and the hole at the same site, and has the excitation energy $2g$.
}
\end{center}
\end{figure*}

We first introduce a 1D toy model of a multiferroic ferromagnet.
The electromagnon excitation can be described by a Hamiltonian for electrons where inversion symmetry is broken, the spin orbit coupling (SOC) is present, and magnetic order is developed with an interaction term.
Specifically we consider the Hamiltonian,
\begin{align}
\mathcal{H} &= H_0 + H_{SOC} + H_{int} + H_{g},
\end{align}
which is schematically illustrated in Fig.~\ref{fig: model}(b).

The first two terms in $\mathcal{H}$ are the single particle particle part of the Hamiltonian.
The first term $H_0$ describes
hopping of electrons in the 1D chain with staggered potential,
and is given by
\begin{align}
H_0 &= \sum_{i,s} [t (c_{i+1,s}^\dagger c_{i,s} + h.c.) 
+ (-1)^i m c_{i,s}^\dagger c_{i,s}].
\end{align}
Here $c_{i,s}$ is the annihilation operator at the site $i$,
$s = \uparrow (+), \downarrow (-)$ indicates the spin degrees of freedom, $t$ is the hopping amplitude, and $m$ is the strength of the staggered potential.
The second term $H_{SOC}$ is the spin orbit coupling. For simplicity, we consider spin orbit coupling proportional to the spin flipping term $s_x$ as
\begin{align}
H_{SOC} &= 
\sum_{i,s, s'} [\lambda + (-1)^i \delta \lambda] [c_{i+1,s}^\dagger (s_x)_{s,s'} c_{i,s'} + h.c.],
\end{align}
where $\lambda$ and $\delta \lambda$ are uniform and alternating parts of the spin orbit interaction, respectively,
and $s_a$ $(a=x,y,z)$ is the Pauli matrix acting on the spin degrees of freedom.
We note that the alternating part ($\delta \lambda$ term) breaks the inversion symmetry.
Since the inversion breaking and SOC are incorporated in the single particle part of $\mathcal{H}$,
one can expect electromagnon excitation once a magnetic order is introduced by the interaction term.

The last two terms in $\mathcal{H}$ are the interaction part of the Hamiltonian.
The third term $H_{int}$ is the onsite Hubbard interaction, given by 
\begin{align}
H_{int} &= U \sum_i \left(n_{i,\uparrow} - \frac 1 2 \right) \left(n_{i, \downarrow} - \frac 1 2 \right),
\end{align}
where $n_{i,s} = c_{i,s}^\dagger c_{i,s}$
 is the density operator at site $i$.
We further introduce the spin anisotropy term
\begin{align}
H_{g} &= - g \sum_m \hat s_{z, 2m} \hat s_{z, 2m+1},
\end{align}
which plays a role in gapping out the electromagnon dispersion and stabilizing the magnetic order.
Here 
$\hat s_{z,i}= \sum_{s,s'} c_{i,s}^\dagger (s_z)_{s s'} c_{i,s'}$
 is the $z$ component of the spin operator at the site $i$, and 
$g$ is the strength of Ising anisotropy along the $z$ direction.
The anisotropy is incorporated with an Ising interaction for pairs of neighboring electron spins within the same unit cell. (Note that the onsite anisotropy $\propto s_{z,i}^2$ reduces to Hubbard interaction and is not able to gap out the electromagnon excitation.)
With this Ising interaction, the ferromagnetic state is realized (at least as a meta-stable state).

\begin{figure}
\begin{center}
\includegraphics[width=0.95\linewidth]{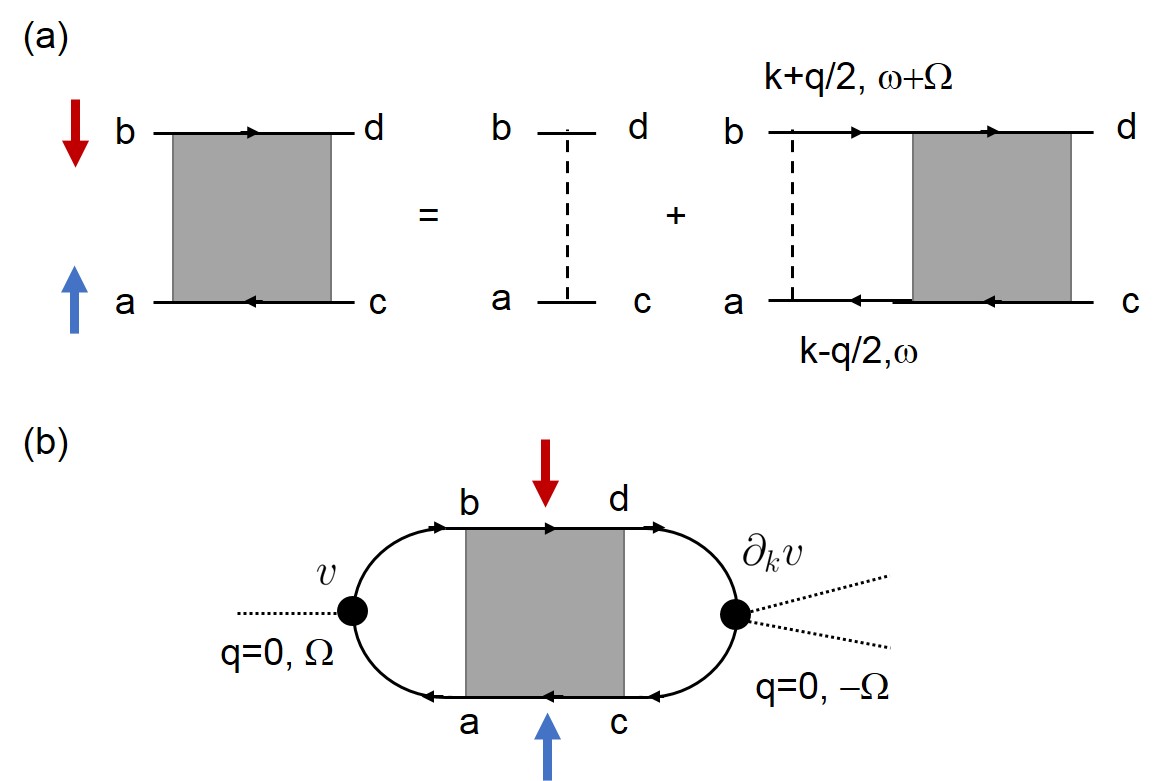}
\caption{\label{fig: diagrams}
(a) Ladder diagram for the magnon propagator.
Solid lines represent electron propagators $G$, and the dashed lines represent the Hubbard interaction $U$. The shaded region is the electron-hole propagator $\Pi$ (i.e., the magnon propagator).
(b) Bubble diagram for the shift current response.
The vertices $v$ and $\partial_k v$ are paramagnetic and diamagnetic current operators, respectively, which couple the electrons with the external electric fields represented by the dotted lines. 
}
\end{center}
\end{figure}

\section{Magnon excitation in the 1D model}
In this section,  we study magnon excitation in the 1D model by ladder approximation for Hubbard interaction.
We neglect the SOC ($\lambda$ and $\delta \lambda$) in this section, since we treat the SOC perturbatively in the shift current response and it turns out that the lowest order contribution in $\lambda$ for the shift current response only involves the $O(\lambda^0)$ term of the magnon propagator in the SOC $\lambda$.

We consider a groundstate with a ferromagnetic order within the mean field approximation, where the up spin is fully polarized ($\braket{n_{i,\uparrow}}=1$ and $\braket{n_{i,\downarrow}}=0$).
The mean field Hamiltonians for the up and down spin states are given by $H_0 - (U/2 +g)$ and $H_0 + (U/2 +g)$, respectively.
Thus, the propagators of the electrons are diagonal with respect to spins and are given by
\begin{align}
G_s(k,i\omega) &= \frac{1}{i\omega-H_0+s(\frac{U}{2}+g) }.
\label{eq: G}
\end{align}
in the Matsubara frequency formalism,
where $s=\pm 1$ (or $\uparrow, \downarrow$) denotes the up and down spin states. 
(We set $e=1$ and $\hbar=1$ in this section, for simplicity.)
We note that the Green's function $G_s$ is a two-dimensional matrix spanned over the sublattice degrees of freedom, 
\begin{align}
\{ \ket{A}, \ket{B} \}.
\end{align}

The Hubbard interaction induces pairing between the electrons and holes, leading to the exciton formation (Fig.~\ref{fig: model}(c)).
The spin wave can be regarded as a triplet exciton, and hence,
the (electro)magnon excitation is described by the propagator of electron hole-pairs in the spin triplet states as shown in Fig.~\ref{fig: diagrams}(a).
For triplet states, we adopt the four dimensional base, 
\begin{align}
\{
\ket{A \uparrow A\downarrow}, \ket{A \uparrow B\downarrow}, \ket{B \uparrow A\downarrow}, \ket{B \uparrow B\downarrow} 
\}
\label{eq: base 4}
\end{align}
to represent the electron-hole propagators (i.e., magnon propagators) $\Pi_0$ in the matrix form, where the first (second) component $A/B\uparrow(\downarrow)$ of the basis element specifies the sublattice degrees of freedom for the hole (electron) state with the up (down) spin. 
Using the Green's function in Eq.~\eqref{eq: G}, we can write the bare electron-hole propagator as 
\begin{align}
&(\Pi_0)_{cd; ab}(q, i\Omega) \n
&= 
\mathfrak{a} \int \frac{d\omega}{2\pi}\frac{dk}{2\pi}
G_{\uparrow, a c}(k-\frac{q}{2}, i\omega)
G_{\downarrow, d b}(k+\frac{q}{2}, i\omega+i\Omega),
\end{align}
where $q$ and $\Omega$ are the momentum and the frequency of the electron-hole pair, respectively, and $\mathfrak{a}$ is the lattice constant.
Here the subscripts $a,b,c,d$ are sublattice indices running $A, B$; specifically, $a,c$ correspond to the sublattice indices for holes with the up spin, and $b,d$ for electrons with the down spin (Fig.~\ref{fig: diagrams}(a)).
The effective interaction between the electrons and the holes (the four vertex function) $\Gamma$ is given in the ladder approximation by
\begin{align}
\Gamma = - \hat U - \hat U \Pi_0 \Gamma
\end{align}
as depicted in Fig.~\ref{fig: diagrams}(a).
This leads to the expression,
\begin{align}
\Gamma &= - (1+ \hat U \Pi_0)^{-1} \hat U.
\end{align}
Here we used the onsite Hubbard interaction represented in the four by four diagonal matrix form, 
\begin{align}
\hat U = U 
\begin{pmatrix}
1 &&& \\
& 0&& \\
&& 0& \\
&&&1
\end{pmatrix},
\end{align}
which means that the an effective interaction works only for $\ket{A \uparrow A\downarrow}$ and $\ket{B \uparrow B\downarrow}$.
The magnon propagator $\Pi$ is obtained from the vertex function $\Gamma$ via 
\begin{align}
\Pi = \Pi_0 + \Pi_0 \Gamma \Pi_0,
\end{align}
indicating that $\Pi$ and $\Gamma$ share the same pole structures for low energy excitations such as the magnon excitations.
Thus, we look at the pole structure of $\Gamma$ to study the magnon dispersion in the following.
We note that we do not incorporate the effect of Ising anisotropy $H_g$ in the ladder summation for $\Gamma$, while we do so for the electron Green's function $G(\omega)$ within the mean field approximation. This is because $H_g$ only gives rise to a pole in $\Gamma(\omega)$ at high energy region ($\omega \sim U $) and does not change the pole structure at the lower energy region including the magnon excitations around $\omega \sim 2g $.

\begin{figure}
\begin{center}
\includegraphics[width=0.75\linewidth]{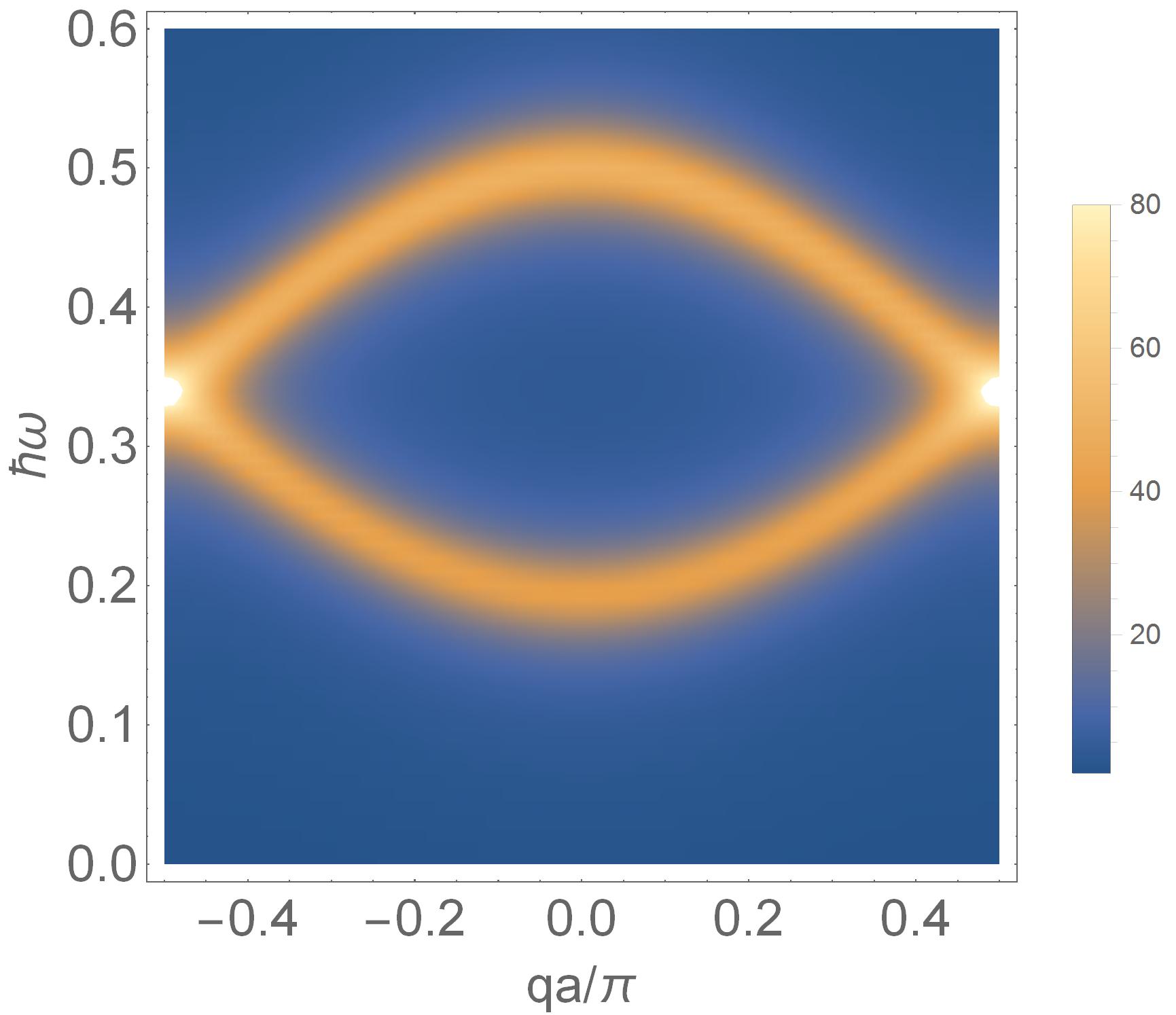}
\caption{\label{fig: magnon dispersion}
The spectral weight of the four vertex function $\t{Im}\{\t{tr}[\Gamma(q,\omega)]\}$ in the 1D model of ferromagnets.
The bright region shows the magnon band dispersion relationship. We used the parameters $t=0.25, m=0.1, U=1.5, g=0.25, \gamma=0.03$ (the units are eV).
The magnon dispersion in the upper branch appears at $\hbar\omega=2g$ at $q=0$, and the magnon energy decreases for $q \neq 0$.
The lower branch appears from the band folding due to the staggered potential.
}
\end{center}
\end{figure}

Figure~\ref{fig: magnon dispersion} shows the spectral weight of the four vertex function $\t{Im}\{\t{tr}[\Gamma(q,\omega)]\}$, which indicates the magnon dispersion.
Specifically, the magnon modes correspond to the eigenvectors of the 4 by 4 matrix $\Gamma(q,\omega)$, and the poles of the corresponding eigenvalues give the magnon dispersion relationship, which can be visualized by plotting the sum of the imaginary parts of the eigenvalues ($=\t{Im}\{\t{tr}[\Gamma(q,\omega)]\}$).
We used the parameters $t=0.25$, $m=0.1$, $U=1.5$, $g=0.25$, and performed analytic continuation of the Matsubara frequency $i\Omega \to \omega + i \gamma$ where $\gamma=0.03$ gives the energy broadening for the magnon dispersion.
The bright region in Fig.~\ref{fig: magnon dispersion} indicates the magnon dispersion relationship.
There appear two magnon bands 
since the noninteracting part of the electron Hamiltonian is a two band system due to the staggered potential.
The upper branch of the magnon excitation appears at $\omega=2g$ for $q=0$ since the Ising anisotropy introduces a gap for the magnon dispersion.
As a function the momentum $q$, the magnon energy decreases due to the effect of the hopping $t$. 
The lower magnon branch appears from the band folding due to the staggered potential,
and the magnon in the lower branch at $q=0$ is folded from $q=\pi$ state which shows the largest energy decrease due to the hopping of electrons.
Also, the energy gap for the magnon excitation indicates that the ferromagnetic ground state is (at least locally) stable.

If we focus on the case $q=0$ which is relevant for photoexcitations,
we can derive an analytic expression for $\Gamma(q=0, i\Omega)$, by treating the hopping term perturbatively when $t \ll m ,g, U$. 
When we are interested in the lowest order contribution in $t$ to the shift current response,
we can neglect the effect of $t$ in the electron-hole propagator $\Pi_0$,
since the hopping term $\propto t$ appears at the vertex of photoexcitation in the diagram for nonlinear conductivity (Fig.~\ref{fig: diagrams}(b)), as we explain in detail later.
Within this perturbative treatment, the bare electron-hole propagator (which we denote by $\Pi_0^\t{atomic}$)  reduces to a 4 by 4 diagonal matrix,
\begin{align}
&[\Pi_0^\t{atomic} (q=0, i\Omega)]^{-1} \n
&=
i\Omega- (U+2g) -
\begin{pmatrix}
0 &&& \\
& -2m  && \\
&& 2m & \\
&&&0 
\end{pmatrix},
\end{align}
in the basis in Eq.~\eqref{eq: base 4}.
After summing the ladder diagram, the four vertex function $\Gamma^\t{atomic}$ (within this perturbative treatment) is obtained as
\begin{align}
&\Gamma^\t{atomic}(q=0, i\Omega) \n
&= U(U+2g-i\Omega)
\begin{pmatrix}
\frac{1}{i\Omega-2g} &&& \\
& 0  && \\
&& 0 & \\
&&& \frac{1}{i\Omega-2 g} 
\end{pmatrix}.
\end{align}
This expression shows that the magnon excitations are formed by $\ket{A\uparrow A \downarrow}$ and $\ket{B\uparrow B \downarrow}$ through Hubbard interaction and the magnon excitation energy is given by $i\Omega=2g$ as expected (Fig.~\ref{fig: model}(c)).
This approximation gives a consistent result with the full calculation for $\Gamma$ shown in Fig.~\ref{fig: magnon dispersion}.
We note that the two magnon modes in Fig.~\ref{fig: magnon dispersion} happen to be degenerate in this perturbative treatment, which is an artifact of neglecting the hopping $t$ in the propagator.
We also show a brief derivation of the full magnon dispersion in the case of $m=0$ in Appendix~\ref{app: magnon}.

\section{Shift current from electromagnon excitations}
In this section we study shift current of electromagnon by computing the nonlinear conductivity with the bubble diagram.
The SOC plays a crucial role since it provides two necessary ingredients for shift current from the electromagnons, i.e., (i) the coupling of magnon to the external electric field and (ii) the inversion symmetry breaking.
(i) Photoexcitation of electromagnons requires a spin orbit coupling. Since electromagnon is a triplet exciton, its creation involves spin flipping. When the SOC is present via $H_{SOC}$, spin flipping can be induced by an external electric field. 
(ii) In addition, $H_{SOC}$ breaks the inversion symmetry with the alternating hopping ($\delta \lambda$ term). Inversion breaking leads to nonzero polarization of triplet exciton states, and enables dc current response through their photoexcitations.

We study the nonlinear current response, 
\begin{align}
J_{dc} = \sigma^{(2)}(\omega) E(\omega) E(-\omega),
\end{align}
where dc current $J_{dc}$ is induced by irradiating light of the frequency $\omega$.
In the shift current mechanism, the nonlinear conductivity $\sigma^{(2)}(\omega)$ is given by a correlation function of paramagnetic and diamagnetic current operators (Fig.~\ref{fig: diagrams}(b))~\cite{Morimoto-Nagaosa16,Nagaosa-Morimoto17,Kim17}.
Namely, we can write the nonlinear conductivity as
\begin{align}
\sigma^{(2)}(i\Omega) &= \frac{2\pi e^3 \mathfrak{a}}{\hbar^2 \omega^2} \sum_{abcd} 
 V^D_{cd}(-i\Omega) \Gamma_{cd;ab} (q=0,i\Omega) V^P_{ba}(i\Omega) ,
\end{align}
where $V^P$ and $V^D$ are the paramagnetic and diamagnetic current vertices, respectively, that are represented by two by two matrices spanned over the sublattice degrees of freedom ($\ket{A}$ and $\ket{B}$) which we label by the subscripts ($a, b, c, d$). 
To define the two current vertices, we first define the full electron Green's function $\tilde{G}$ and the full current operator $\tilde{v}$ that include the effect of the SOC as
\begin{align}
\tilde{G}(k, i\omega) &= [i\omega - (H_0(k) + H_{SOC}(k))]^{-1}, \\
\tilde{v}(k) &= \frac 1 \hbar \partial_k (H_0(k) + H_{SOC}(k)).
\end{align}
Here $\tilde{G}$ and $\tilde{v}$ are 4 by 4 matrices spanned over the four states with the sublattice and spin degrees of freedom: 
\begin{align}
\{ \ket{A\uparrow}, \ket{B\uparrow}, \ket{A\downarrow}, \ket{B\downarrow} \}.
\end{align}
With these operators including the SOC, the two full current vertices $\tilde V^P$ and $\tilde V^D$ can be written in the 4 by 4 matrix form as
\begin{align}
\tilde V^P(i\Omega) &= \int \frac{d\omega}{2\pi} \frac{dk}{2\pi} \tilde G(k,i\omega+i\Omega) \tilde{v}(k) \tilde G(k,i\omega), \\
\tilde V^D(i\Omega) &= \int \frac{d\omega}{2\pi} \frac{dk}{2\pi} \tilde G(k,i\omega+i\Omega) \partial_k \tilde{v}(k) \tilde G(k,i\omega).
\end{align}
The current vertex $V^P$ in the 2 by 2 form is obtained from $\tilde V^P$ by projecting the spin state of the incoming electron into $\uparrow$ spin and projecting the spin state of the outgoing electron in to $\downarrow$ spin: 
\begin{align}
(V^P)_{ba} &= \bra{b\downarrow} \tilde V^P \ket{a \uparrow}.
\end{align}
Similarly, the 2 by 2 matrix $V^D$ is obtained from $\tilde V^D$ by projection as
\begin{align}
(V^D)_{cd} &= \bra{c\uparrow} \tilde V^D \ket{d \downarrow}.
\end{align}

We note that we use the four vertex function $\Gamma$ that does not include the effect of the SOC for evaluating the nonlinear conductivity $\sigma^{(2)}$. This is an approximation that is justified as far as the SOC $\lambda$ is small compared to other energy scale, which is usually true. Specifically, the matrix elements of the two current vertices $V^P$ and $V^D$ are $O(\lambda)$ in the perturbative expansion in the SOC $\lambda$, and it suffices to only keep the leading order $O(\lambda^0)$ term for the four vertex function $\Gamma$, when we consider the leading order contribution to $\sigma^{(2)}$.

\begin{figure}
\begin{center}
\includegraphics[width=0.8\linewidth]{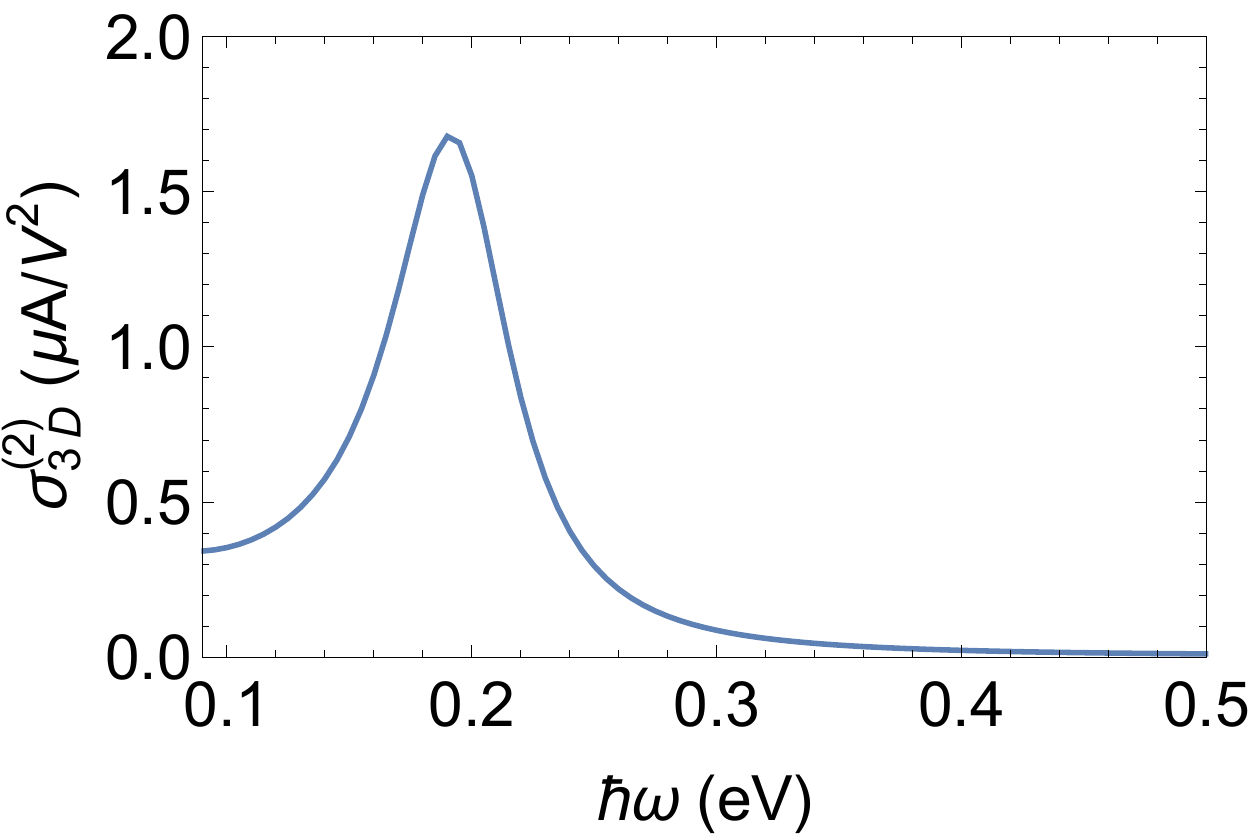}
\caption{\label{fig: sigma 2}
Nonlinear conductivity $\sigma^{(2)}_{3D}(\omega)$. We used the parameters $t=0.25$, $m=0.1$, $U=1.5$, $g=0.25$, $\lambda=0.05$, $\delta \lambda=0.02$, $\gamma=0.03$ (the units are eV).
The peak around $\hbar\omega \simeq 0.2$eV corresponds to the electromagnon excitation at $q=0$, and indicates that the electromagnon excitation supports nonvanishing dc current response. 
}
\end{center}
\end{figure}

We show the numerical result for the nonlinear conductivity $\sigma^{(2)}(\omega)$ in Fig.~\ref{fig: sigma 2}. 
Note that we do not employ perturbative expansion here.
We calculate the two current vertices $V^P$ and $V^P$ with the mean-field Hamiltonian including the SOC, and use the four vertex function $\Gamma$ that we obtained in the previous section.
To facilitate comparison with 3D bulk materials, we look at the 3D nonlinear conductivity 
\begin{align}
\sigma^{(2)}_{3D}(\omega) &\equiv \frac{\sigma^{(2)}(\omega)}{\mathfrak{a}^2}, 
\end{align}
 with the lattice constant $\mathfrak{a}$.
The resulting nonlinear conductivity $\sigma^{(2)}_{3D}$ is plotted in Fig.~\ref{fig: sigma 2}.
We find a peak of $\sigma^{(2)}_{3D}$ around $\omega=0.2$ which corresponds to the magnon excitation at $q=0$ in Fig.~\ref{fig: magnon dispersion}. There does not appear a peak corresponding to the higher energy magnon excitation around $\omega=0.5$ due to a matrix element effect in the bubble diagram.
This numerical result clearly demonstrates that the electromagnon excitation supports nonvanishing dc photocurrent.

We can also derive an analytic expression for $\sigma^{(2)}$ by performing a perturbation theory for the current vertices with respect to the hopping $t$ and SOC $\lambda$.
The magnon propagator has weights only on $\ket{A\uparrow A\downarrow}$ and $\ket{B\uparrow B\downarrow}$,
and the corresponding current vertices arise from combining one hopping perturbation and one SOC perturbation, so that they creates an electron-hole pair in the triplet spin state on the same sublattice.
For example, one can use the SOC to create a $\ket{A\uparrow B\downarrow}$ state and uses the hopping term for $\downarrow$ to convert it into $\ket{A\uparrow A\downarrow}$.
Thus the lowest contributions to the current vertices are given by the second order terms with respect to $t, \lambda,\delta \lambda$.
We show the explicit expressions of $V^P$ and $V^D$ in the lowest order ($O(\lambda t)$) in Appendix \ref{app: vertices}.
This allows us to compute the nonlinear conductivity analytically (as detailed in Appendix~\ref{app: vertices}) and obtain
\begin{align}
\sigma_{3D}^{(2)}(\omega) &= 
 \frac{4\pi^2e^3}{\hbar^2} \frac{ m \lambda \delta \lambda t^2}{\omega^5}
\delta \left(\omega-\frac{2g}{\hbar} \right).
\end{align}
This expression clearly shows that the electromagnon can be excited with light irradiation at $\hbar\omega=2g$ from the SOC and induces shift current response due to the inversion breaking introduced with $\delta \lambda$.
Here two degenerate magnon modes contribute to the shift current as detailed in Appendix~\ref{app: vertices}.

It has been known that electromagnons can be excited by light irradiation, typically in the THz regime, and its photoexcitation accompanies electric polarization. 
Our calculation of $\sigma^{(2)}(\omega)$ demonstrates that photoexcitation of electromagnon also leads to dc photocurrent.
This phenomenon can be understood from the fact that creation of one electromagnon accompanies electric polarization characterized by the real-space shift of the wave packets between the spin up and the spin down states.
Constant irradiation of light induces growing density of electromagnons, and hence, the electric polarization grows linearly in time. 
Since the time derivative of polarization gives dc current, 
such constant creation of electromagnons naturally leads to constant dc current.
In short, steady dc current induced by spin excitation arises from the multiferroic nature of the system.
Therefore shift current of multiferroics will provide a novel way to study multiferroic materials through optical measurements.

\section{Discussions}
As discussed for the shift current due to electron-hole excitations, the decay processes are crucial for the dc current. Namely, the recombination of the electron and the hole leads to the cancellation of the shift current via the inverse process of the excitation. Similar situation applies here; the electromagnon decay by emitting light cancels the shift current. Therefore, the charge carriers must be created at the contact with the electrodes to avoid complete cancellation of the dc current. 
Alternatively, the time-domain experiment would be suitable to demonstrate the existence of the shift current due to electromagnon \cite{Sotome18,Sotome-apl}. 
For example, Mn perovskites support electromagnons in the THz region \cite{Kida}, and would be suitable for the time-domain experiment.

The magnitude of the shift current from the electromagnon excitations should be comparable to the usual shift current induced by electron hole pair creation over the band gap.
This is because electromagnon can be regarded as a triplet exciton and the exciton supports the polarization comparable to that of free electron-hole pairs induced by interband optical transition. As far as there exists sufficient SOC to support photoexcitation of electromagnons, the magnitude of the resulting shift current is comparable to the conventional one.
In addition, since the shift current is enhanced in low frequency, shift current of electromagnon can be significantly enhanced by this low frequency factor.
For example, the nonlinear conductivity $\sigma_{3D}^{(2)} (\omega)$ in Fig.~\ref{fig: sigma 2} is typically in the order of $\mu A/V^2$. 
If we consider an electric field of the light $E\simeq 100V/cm$ in the (far) infrared regime where electromagnon excitation appears, 
the current density $\sigma_{3D}^{(2)} (\omega) E^2$ becomes $j\simeq 10^2 A/m^2$. For a sample of the cross section 1mm$^2$, the current amounts to $J= 0.1A$, which is large enough for experimental detection. 

\begin{acknowledgements}
We thank Daniel I. Khomskii and Yoshinori Tokura for insightful discussions.
This work was supported by The University of Tokyo Excellent Young Researcher Program, JST PRESTO (JPMJPR19L9), JST CREST (JPMJCR19T3)(TM), and
JST CREST (JPMJCR1874 and JPMJCR16F1), 
JSPS KAKENHI (18H03676 and 26103006)(NN).
\end{acknowledgements}

\appendix
\section{Ferromagnetic phase in 1D Hubbard model \label{app: magnon}}
We study the ferromagnetic phase in 1D Hubbard model.
We consider the mean-field Hamiltonian,
\begin{align}
H=
\begin{pmatrix}
c^\dagger_\uparrow & c^\dagger_\downarrow
\end{pmatrix}
\begin{pmatrix}
\frac U 2 + \epsilon_k & 0 \\
0 & - \frac U 2 + \epsilon_k
\end{pmatrix} 
\begin{pmatrix}
c_\uparrow \\ c_\downarrow
\end{pmatrix}
+ U n_\uparrow n_\downarrow.
\end{align}
We compute the ladder diagram as
\begin{align}
&\int d\omega [dk] G_\uparrow(k-\frac q 2, i \omega) G_\downarrow (k+\frac q 2, i \omega + i\Omega)  \n
&= 
\int [dk] \frac{1}{i\Omega + \epsilon_{k-\frac q 2} - \epsilon_{k+\frac q 2}- U},
\end{align}
with $[dk]=\frac{dk}{2\pi}$.
Analytic continuing $i\Omega \to \omega$ and taking the ladder summation gives
\begin{align}
\chi(q,\omega) &=
\frac{U}
{1+U
\int [dk] \frac{1}{\omega + \epsilon_{k-\frac q 2} - \epsilon_{k+\frac q 2}- U}.
}.
\end{align}
If we consider the dispersion $\epsilon_k= \cos k$,
we can compute the integral as
\begin{align}
\int [dk] \frac{1}{\omega + \epsilon_{k-\frac q 2} - \epsilon_{k+\frac q 2}- U}
&=
- \frac{1}{\sqrt{(\omega-U)^2 -2 +2 \cos q}}.
\end{align}
The pole of the ladder diagram is obtained from
\begin{align}
\frac{1}{U}=\frac{1}{\sqrt{(\omega-U)^2 -2 +2 \cos q}},
\end{align}
which leads to the magnon dispersion,
\begin{align}
\omega=U \pm \sqrt{U^2 + 4 \sin^2 q}.
\end{align}
The lower branch shows $\omega(q=0)=0$ as expected from the existence of Goldstone mode, but $\omega(q\neq 0)$ becomes negative indicating instability towards antiferromagnetic phase.

Introducing Ising spin anisotropy stabilizes the ferromagnetic phase and introduces a gap to the magnon band dispersion.
By introducing $H_g$ and treating it in the mean field approximation, the magnon dispersion $E(q)$ is shifted as $E(q) \to E(q)+2g$.
Thus the lower branch of the magnon dispersion is given by
\begin{align}
E(q) &= 2g + U - \sqrt{U^2 + 4t^2 \sin^2 q} \n
&= 2g - \frac{2t^2}{U} \sin^2 q + O\left(\frac{1}{U^2}\right). 
\end{align}

\section{Expressions of current vertices from the second order perturbation theory \label{app: vertices}}
In this section, we give perturbative expressions for the current vertices and derive an analytic expression for the bubble diagram for $\sigma^{(2)}$. We set $e=1, \hbar=1, \mathfrak{a}=1$ in this section, for simplicity.

We treat the hopping $t$ and the SOC $\lambda, \delta \lambda$ are treated perturbatively, and define the electron Green's function in the atomic limit $G_{0,s}$ (setting $t=0$ in $G_s$) as
\begin{align}
G_{0,s}(k, i\omega) &= \left[i\omega - \hat m + s \left(\frac{U}{2}+g \right) \right]^{-1},
\end{align}
with $s=\pm$ being the spin and $\hat m \equiv m \sigma_z$,
where $\sigma_z$ is the Pauli matrix acting on the sublattice degrees of freedom.
The current vertices are then obtained from the second order perturbation with $t$ and $\lambda, \delta \lambda$ as
\begin{widetext}
\begin{align}
V^P(i\Omega) \simeq \int d\omega dk 
G_{0,\downarrow}(k,i\omega+i\Omega) &[
h_{soc}(k) G_{0,\uparrow}(k,i\omega+i\Omega) \hat{v_t}(k)  
 + h_t(k) G_{0,\downarrow}(k,i\omega+i\Omega) \hat{v}_{soc}(k)  \n
& + \hat{v}_{soc}(k) G_{0,\uparrow}(k,i\omega) h_t(k)  
 + \hat{v_t}(k) G_{0,\uparrow}(k,i\omega) h_{soc}(k)
]G_{0,\uparrow}(k,i\omega), 
\end{align}
and
\begin{align}
V^D(-i\Omega) \simeq \int d\omega dk 
G_{0,\uparrow}(k,i\omega) 
&[
h_{soc}(k) G_{0,\downarrow}(k,i\omega) \partial_k \hat{v_t}(k)  
 + h_t(k) G_{0,\uparrow}(k,i\omega) \partial_k \hat{v}_{soc}(k)  \n
& + \partial_k \hat{v}_{soc}(k) G_{0,\downarrow}(k,i\omega+i\Omega) h_t(k)  
 + \partial_k \hat{v_t}(k) G_{0,\uparrow}(k,i\omega+i\Omega) h_{soc}(k)
]G_{0,\downarrow}(k,i\omega+i\Omega). 
\end{align}
Here we defined the hopping perturbation term $h_t(k) = t\cos k \sigma_x $, the SOC perturbation term $h_{soc}(k) = \lambda \cos k \sigma_x  + \delta \lambda \sin k \sigma_y$, and their corresponding velocity operators $\hat{v}_{t/soc} = \partial_k h_{t/soc}(k)$.

After the $\omega$ and $k$ integral, we obtain
\begin{align}
V^P(i\Omega) &=
-i \delta\lambda t \sigma_z 
\frac{ 4 \hat m^2 (U-3 i\Omega)-4 \hat m (U-i\Omega)^2+U^3-3 U^2 i\Omega+U i\Omega^2+i\Omega^3 }
{2 (2 \hat m-U) (2 \hat m+i\Omega) (U-i\Omega) (2 \hat m+U-i\Omega) (2 \hat m-U+i\Omega)}
\end{align}
and
\begin{align}
V^D(-i\Omega)
&=
\lambda t 
\frac{32 \hat m^3-12 \hat m^2 (U-i\Omega)-4 \hat m (U^2-U i\Omega+i\Omega^2)+(U-i\Omega)^3}{2 (2 \hat m-U) (2 \hat m+i\Omega) (U-i\Omega) (2 \hat m+U-i\Omega) (2 \hat m-U+i\Omega)}.
\end{align}
\end{widetext}
When the Hubbard interaction is large enough, we can expand the above expressions with respect to $1/U$ as
\begin{align}
V^P(i\Omega) &= 
-i \sigma_z \frac{\delta\lambda t}{2 U (2\hat m + i\Omega)} + O\left(\frac{1}{U^2}\right), \\
V^D(-i\Omega) &=
\frac{\lambda t}{2 U (2\hat m + i\Omega)} + O\left(\frac{1}{U^2}\right).
\end{align}
Using these, we can write the summand in the expression for the nonlinear conductivity as
\begin{align}
&\sum_{abcd} V^D_{cd}(-\omega) \Gamma_{cd;ab}(q=0,\omega) V^P_{ba}(\omega) \n
&= \sum_{ab} V^D_{ab}(-\omega) V^P_{ba}(\omega) [-i\pi U^2 \delta(\omega-2g)] \n
&= -i \t{tr} \left[ \sigma_z \frac{\lambda \delta\lambda t^2}{[2 U (2 m \sigma_z + \omega)]^2} [- i\pi U^2 \delta(\omega -2g)] \right] \n
&= 2 \pi \frac{m \lambda \delta\lambda t^2}{\omega^3} \delta(\omega -2g).
\end{align}
in the lowest order contribution with respect to $t,\lambda, \delta \lambda, m$.
This leads to a perturbative expression for $\sigma^{(2)}_{3D}$ as
\begin{align}
\sigma^{(2)}_{3D} &= 
\frac{4\pi^2e^3}{\hbar^2} \frac{ m \lambda \delta \lambda t^2}{\omega^5} \delta \left(\omega -\frac{2g}{\hbar}\right).
\end{align}
Here, we recovered $e$ and $\hbar$.
This expression indicates that the shift current appears at the electromagnon excitation at $\hbar\omega= 2g$ and is proportional to the strength of inversion breaking $\delta \lambda$.

\bibliography{EM}

\end{document}